\documentclass[twocolumn,showpacs,showkeys,preprintnumbers,
amsmath,amssymb,aps,floatfix,pre]{revtex4}
 
\usepackage{graphicx}
\usepackage{epstopdf}
\usepackage{amsmath}
\usepackage{multirow}
\usepackage{times}
\usepackage{amssymb}
\usepackage{dcolumn}%
\usepackage{bm}%
\usepackage[dvipdfmx]{hyperref}%
\usepackage{upgreek}
\usepackage{wasysym}

\hypersetup
{
	colorlinks=true,
	linkcolor=blue,
	citecolor=blue
}

\begin{document}

\title[]{Cage effect in supercooled molecular liquids: local anisotropies and collective solid-like response}
\author{S.\@ Bernini \footnote{Present 
 address: Jawaharlal Nehru Center for Advanced Scientific Research, Theoretical Sciences Unit, Jakkur Campus, Bengaluru 560064, India.}}
\affiliation{Dipartimento di Fisica ``Enrico Fermi'', 
Universit\`a di Pisa, Largo B.\@Pontecorvo 3, I-56127 Pisa, Italy}
\author{D.\@ Leporini}
\email{dino.leporini@unipi.it}
\affiliation{Dipartimento di Fisica ``Enrico Fermi'', 
Universit\`a di Pisa, Largo B.\@Pontecorvo 3, I-56127 Pisa, Italy}
\affiliation{IPCF-CNR, UOS Pisa, Italy}

\date{\today}

\begin{abstract} 
Both local geometry and collective, extended excitations drive the moves of a 
particle in the cage of its neighbours in dense liquids. The strength of their 
influence is investigated by Molecular Dynamics simulations of a supercooled 
liquid of fully-flexible trimers with semirigid or rigid bonds. The rattling in 
the cage is investigated on different length scales. First,  the rattling 
anisotropy due to {\it local order} is characterized by two order parameters 
sensing the monomers succeeding or failing  to escape from the cage. Then,  the 
collective response of the surroundings excited by the monomer-monomer 
collisions is considered. The collective response is initially restricted to 
the nearest neighbours of the colliding particle by a Voronoi analysis 
revealing elastic contributions.  Then, the {\it long-range} excitation of the 
farthest neighbours is scrutinised  by searching spatially-extended 
correlations between the simultaneous fast  displacements of the caged particle 
and the surroundings. It is found that the longitudinal component has stronger 
spatial modulation than the transverse one with wavelength of about one 
particle diameter, in close resemblance with experimental findings on colloids.
 It is concluded that the cage rattling is largely affected by solid-like 
extended modes.
  \end{abstract}

\maketitle

\section{Introduction}
\label{intro}

The relation between the structure and the dynamics is a key problem in liquid-state 
\cite{HansenMcDonaldIIIEd,GotzeBook,BerthierBiroliRMP11,EdigerHarrowell12} and polymer 
\cite{PaulSmithRepProgrPhys04,sim} physics.  Here, we address the case of dense supercooled liquids  where each particle 
is temporarily trapped by the cage formed by the first neighbours. The lifetime of the cage is set by the structural 
relaxation and quantified by the structural relaxation time $\tau_\alpha$ which exceeds the rattling times of the 
particle in the cage of several orders of magnitude on approaching the glass transition from above. 

Which are the key aspects driving the moves of the trapped particle in the cage 
? At very short times, fractions of picoseconds, the {\it local geometry} plays 
the leading role. This inspired the free-volume model \cite{Debenedetti97} 
which has been recently re-examined 
\cite{SastryEtAl98,StarrEtAl02,DouglasCiceroneSoftMatter12,DouglasAspehricity}. 
The role of the local structure is seen by e.g. considering the short-time 
expansion of the mean square displacement (MSD):
\begin{equation}
\langle r^2(t) \rangle = 3 \, \upsilon^2 t^2 -  \frac{1}{4} \, \upsilon^2 \, \Omega_0^2 \, t^4  + \cdots
\label{MSDexpansion}
\end{equation}
where $\upsilon = \sqrt{k_B T/m}$ is the thermal velocity \cite{Boon}. Initially,  MSD is ballistic but early collisions with the first neighbours slow down the particle. $\Omega_0$ is an effective collision frequency. More precisely, $\Omega_0$ is the frequency at which the tagged particle would vibrate if it were undergoing small oscillations in the potential well produced by the surrounding monomers when kept {\it fixed} at their mean equilibrium positions \cite{Boon}. Collisions lead also to correlation loss of the velocity and the related correlation function starts to decay as
\begin{equation}
 C_{vv}(t) =  3 \upsilon^2 \left( 1 - \frac{\Omega_0^2}{2} t^2 + \cdots \right)
\label{Cvvexpansion}
\end{equation}
 After a few collisions, velocity correlations reveal oscillatory components due to  sound waves that, owing to the low compressibility of liquids, reach wavelengths of a few particle diameters \cite{HansenMcDonaldIIIEd,GaskellMillerJPhysC78}. This means that the particle displacement is also affected by {\it collective, elastic modes}. Later, the particle completes the exploration of the cage in a time $t^*$, a few picoseconds, and, in the absence of escape processes, MSD would levels off at $\langle r^2(t^*) \rangle \equiv \langle u^2 \rangle$, the mean square amplitude of the cage rattling (related to the Debye-Waller factor). In actual cases,  early breakouts cause MSD to increase for $t > t^*$ and an inflection point appears at $t^*$ in the log-log plot of $\langle r^2(t) \rangle$ \cite{GotzeBook,StarrEtAl02,SastryPRL16}.

We aim at clarifying by extensive molecular-dynamics (MD) numerical simulations 
of a supercooled molecular liquid if the single-particle fast dynamics up to $ 
\sim t^*$ is more contributed by {\it the local geometry of the cage or the {\it 
solid-like} extended modes}.  To be more precise, the influence of the local 
geometry will be examined by considering how the {\it positions} of the 
particles forming the cage at a given initial time affect the direction of the 
subsequent displacement of the particle in the cage. Instead,  the influence of 
extended collective modes will be studied by 
 the correlations between the direction of the  displacement of the particle 
in the cage with the {\it simultaneous displacements} of the surrounding 
particles.
The present study contributes to our continuing effort to understand the 
microscopic origin of the universal correlation between the fast dynamics, by 
using  $\langle u^2 \rangle$ as metric, and the relaxation and transport, found 
in 
simulations of polymers \cite{OurNatPhys,lepoJCP09,Puosi11}, supercooled binary atomic mixtures \cite{lepoJCP09,SpecialIssueJCP13}, 
colloidal gels \cite{UnivSoftMatter11} and antiplasticized polymers 
\cite{DouglasCiceroneSoftMatter12,DouglasStarrPNAS2015},
and supported by the experimental data concerning several glassformers in  a wide fragility range ($20 \le m \le 191$) 
\cite{OurNatPhys,UnivPhilMag11,OttochianLepoJNCS11,SpecialIssueJCP13,CommentSoftMatter13}.  
On a wider perspective, our investigations on the correlation between $\langle 
u^2 \rangle$ and the relaxation are part of the intense ongoing research on the 
relation between the vibrational dynamics and the relaxation in glassfoming 
systems. As a matter of fact, despite the huge range of time scales older 
\cite{TobolskyEtAl43} and recent theoretical 
\cite{Angell95,HallWoly87,Dyre96,MarAngell01,Ngai04,Ngai00,
DouglasCiceroneSoftMatter12,WyartPRL10} studies addressed the  rattling process 
in the cage to understand the structural relaxation, gaining support from 
numerical 
\cite{Angell68,Nemilov68,Angell95B,StarrEtAl02,BordatNgai04,Harrowell06,
Harrowell_NP08,HarrowellJCP09,DouglasEtAlPNAS2009,XiaWolynes00,DudowiczEtAl08,
DouglasCiceroneSoftMatter12,DouglasStarrPNAS2015,OurNatPhys,lepoJCP09,Puosi11,
SpecialIssueJCP13,UnivSoftMatter11,CommentSoftMatter13,OttochianLepoJNCS11,
UnivPhilMag11,Puosi12SE,Puosi12,PuosiLepoJCPCor12,PuosiLepoJCPCor12_Erratum,
SastryPRL16} and experimental works on glassforming liquids 
\cite{BuchZorn92,AndreozziEtAl98,DouglasCiceroneSoftMatter12} and glasses 
\cite{MarAngell01,ScopignoEtAl03,SokolPRL,Buchenau04,SokolovNovikovPRL13,
NovikovEtAl05,Johari06}.

The coupling between the 
rattling process and extended, fast modes has been indicated 
\cite{Harrowell_NP08,HarrowellJCP09,SastryPRL16,WyartPRL10}.
Recent support to 
the {\it collective} character of the cage rattling is the  evidence of  spatially extended correlations (up to about the 
fourth shell)  between the simultaneous fast displacements of the caged particle and the surrounding ones  
\cite{PuosiLepoJCPCor12,PuosiLepoJCPCor12_Erratum}. They revealed a rather promising feature, i.e. {\it states with identical 
spatial correlations exhibit  {\it equal} mean square amplitude of the cage rattling $\langle u^2 \rangle$ and structural relaxation time $\tau_\alpha$} 
\cite{PuosiLepoJCPCor12,PuosiLepoJCPCor12_Erratum}. 
The role of extended modes in the cage rattling and the relaxation process is 
also suggested by the so-called elastic models, see refs. \cite{Dyre06,Nemilov06} for excellent reviews  and refs. 
\cite{LemaitrePRL14,GranatoFragilityElasticityJNCS02,WyartPNAS2013,NovikovEtAl05,NoviSoko04,TrachenkoBrazhkinElasticJPCM09,TrachenkoBrazhkinCollecJPCB14Review,SchweizerElastic1JCP14,SchweizerElastic2JCP14,DouglasStarrPNAS2015,Puosi12,ElasticoEPJE15,DyreWangGpJCP12,
BerniniElasticJPolB15} for recent related papers.  Recent improvements include the finding of the universal correlation between  the cage rattling and the linear elasticity drawn by simulation  \cite{Puosi12} and supported by comparison with the 
experiments  \cite{ElasticoEPJE15}.
 
The influence of {\it local} order on the rattling motion in the cage has been recently considered. The local 
structure was found to correlate poorly with the cage rattling and then structural relaxation in liquids of linear trimers \cite{VoroBinarieJCP15,VoronoiBarcellonaJNCS14} and atomic mixtures  \cite{VoroBinarieJCP15}. 
In particular, it was find that: 
\begin{itemize}
\item physical states with {\it equal} mean square amplitude of the cage rattling $\langle u^2 \rangle$ and structural relaxation time $\tau_\alpha$  have {\it different} distributions of the cage geometries  \cite{VoroBinarieJCP15,VoronoiBarcellonaJNCS14};
\item for a given state of a liquid of linear chains (trimers or decamers), the end and the central monomers, which have {\it different} distributions of the cage geometries,  have {\it equal} $\langle u^2 \rangle$ and structural relaxation time $\tau_\alpha$ \cite{VoronoiBarcellonaJNCS14}. 
\end{itemize}
Notice that the coincidence of $\langle u^2 \rangle$ and $\tau_\alpha$ of two states mirror the coincidence of the self-part of the van Hove function $G_s({\bf r},t^*)$ and $G_s({\bf r},\tau_\alpha)$, respectively 
\cite{Puosi11,SpecialIssueJCP13}.
These findings are fully consistent with Berthier and Jack who concluded that the influence of structure on dynamics is weak on short length scale and becomes much stronger on long length scale \cite{BerthierJackPRE07}.
Several approaches  suggest that structural aspects matter in the dynamics of glassforming systems. This includes the 
Adam-Gibbs derivation of the structural relaxation \cite{AdamGibbs65,DudowiczEtAl08} - built on the thermodynamic notion 
of the configurational entropy \cite{GibbsDiMarzio58} -, the mode- coupling theory \cite{GotzeBook} and extensions 
\cite{SchweizerAnnRev10}, the random first-order transition theory \cite{WolynesRFOT07}, the frustration-based approach 
\cite{TarjusJPCM05}, as well as the so-called  elastic models  
\cite{Dyre06,Nemilov06,StarrEtAl02,LemaitrePRL14,GranatoFragilityElasticityJNCS02,WyartPNAS2013,NovikovEtAl05,NoviSoko04,
SchweizerElastic1JCP14,SchweizerElastic2JCP14,DouglasStarrPNAS2015,Puosi12,DyreWangGpJCP12,ElasticoEPJE15,
BerniniElasticJPolB15}
in that the modulus is set by the  arrangement of the particles in mechanical equilibrium and their mutual interactions 
\cite{Dyre06,Puosi12}. It was concluded that the proper inclusion of many-body static correlations in theories of the 
glass transition appears crucial for the description of the dynamics of fragile glass formers \cite{coslovichPRE11}.
The search of a link between structural ordering and slow dynamics motivated several studies in liquids 
\cite{NapolitanoNatCom12,EdigerDePabloNatMat13,BarbieriGoriniPRE04,ReichmannCoslovichLocalOrderPRL14,RoyallNatCom15}
colloids \cite{StarrWeitz05,TanakaNatMater08,TanakaNatCom12} and polymeric systems 
\cite{StarrWeitz05,DePabloJCP05,GlotzerPRE07,LasoJCP09,BaschnagelEPJE11,MakotoMM11,LariniCrystJPCM05}.

To discriminate between the roles of the local geometry and the 
collective extended modes in the single-particle  vibrational dynamics, the 
cage rattling will be examined on different length scales. First, we 
characterize the rattling process by {\it local} 
anisotropies, namely order parameters which are projections of the direction of the displacement of the central particle 
onto a fixed local axis. We are inspired by a seminal work by Rahman in an atomic liquid \cite{Rahman66}, studying the  directional correlations between the particle displacement of the trapped particle in the cage and the  position of the centroid 
{\bf C} of the  vertices of the associated Voronoi polyhedron (VP). The interest relies on the fact that the VP vertices 
are located close to the voids between the particles and thus mark the weak spots of the cage. It has been shown in 
simulations of atomic liquids \cite{Rahman66} and experiments on granular matter \cite{DouglasAspehricity} that the  
particle initially moves towards the centroid, so that cage rattling and VP geometry are correlated at {very short} 
times. We extend such studies to later times to reveal the sharp crossover to a regime 
where the anisotropic rattling excites the collective response of the surroundings. The collective response is initially restricted to the nearest neighbours of the colliding particle by investigating statics and fluctuations of the VP surface, volume and  asphericity \cite{LocalOrderJCP13}.  Then, the {\it long-range} excitation of the farthest neighbours is evidenced as spatially-extended correlations between the simultaneous fast 
displacements of the caged particle and the surroundings. 

The paper is organized as follows. In Sec.\;\ref{numerical} the molecular 
models and the MD algorithms are presented. The results are discussed in 
Sec.\;\ref{resultsdiscussion}. In particular, Sec.\ref{GenAsp} presents the 
general aspects of the transport and relaxation of interest. The cage rattling 
process is examined on the local, intermediate and large length scales in 
Sec.\ref{collision},  Sec.\ref{CageGeom}, and Sec.\ref{Collision-driven}, 
respectively. 
Finally, the main conclusions are summarized in Sec. 
\ref{conclusions}.

\section{Methods}
\label{numerical}

A coarse-grained model of a melt of $N_c$ linear fully-flexible molecules with 
three monomers per  chain is considered. Full flexibility is ensured by the 
absence of both torsional or bending potentials hindering the bond 
orientations.
The total number of particles is  $N = 2001$. Non-bonded monomers at a distance $r$ interact via a truncated 
Lennard-Jones (LJ) potential $U_{LJ}(r) =\varepsilon \left [ \left (\sigma^*/ r \right)^{12 } - 2\left 
(\sigma^*/r\right)^6 \right]+U_{cut}$ { for $r< r_c=2.5\,\sigma$} and zero otherwise, 
where $\sigma^*=\sqrt[6]{2} \, \sigma$ is the position of the potential minimum with depth $\varepsilon$. The value of 
the constant $U_{cut}$ is chosen to ensure that $U_{LJ}(r)$ is continuous at $r = r_c$. In the case of {\it semirigid} bonds, the bonded monomers interact by a 
potential which is the sum of the LJ potential and the FENE (finitely extended nonlinear elastic) potential 
$ U^{FENE}(r)=-{1/2} \;kR_0^2 \; \ln\left(1-r^2/R_0^2\right)$ where $k$ measures the magnitude of the interaction and 
$R_0$ is the maximum elongation distance \cite{sim,VoronoiBarcellonaJNCS14}. The parameters $k$ and $R_0$
have been set to $30 \, \varepsilon  / \sigma^2 $ and $ 1.5\,\sigma $ respectively \cite{GrestPRA33}. The resulting bond 
length is $b=0.97\sigma$ within a few percent.  All quantities are in reduced units: length in units of $\sigma$, temperature  in units  of $\varepsilon/k_B$ (with $k_B$ the Boltzmann constant) and time $\tau_{MD}$ in units of $\sigma \sqrt{m / \varepsilon}$ where  $m$ is the monomer mass. We set $m = k_B = 1$. { One time unit corresponds to a few picoseconds \cite{BerniniElasticJPolB15} .}
We investigate states with number density $\rho=1.086$ and temperature $T=0.6, 
0.63, 0.7, 0.8, 0.9,1$. States with {\it rigid} bonds having bond length 
$b_{rigid}=0.97$  are also studied with the same density and $T=0.6,0.9$. We 
also considered a crystalline state, with the same density of the other states, 
obtained by {\it spontaneous} crystallization of an {\it equilibrated} liquid 
made of trimers with rigid bonds at $T=0.7$.
Apart from the crystalline state, the average pressure of the other states 
ranges between $P=6.0$ at $T = 0.6$ and $P = 10.2$ at $T=1$ for the semirigid 
system and very similar results for rigid bonds ( $T = 0.6, P = 6.5$). This 
corresponds to a compressibility factor $Z = P/ \rho T \sim 10$, comparable to 
other studies, e.g. Kremer and Grest found $4.84 \le P \le 5.55$ with density 
$\rho = 0.85$ and $T = 1$, corresponding to $Z \sim 6.1$  
\cite{KremerGrestJCP90}.
Periodic boundary conditions are used. $NVT$ ensemble (constant number of 
particles, volume and temperature) has been used for equilibration runs with 
Nos\'e-Hoover thermostat (damping parameter $0.3$), while $NVE$ ensemble 
(constant number of 
particles, volume and energy) has been used for production runs for a given 
state point \cite{allentildesley}. The simulations of systems with semirigid 
bonds are carried out by using LAMMPS molecular dynamics software 
(http://lammps.sandia.gov) \cite{PlimptonLAMMPS}. The equations of motion of 
the system with rigid bonds are integrated  by using a dedicated software 
developed in-house \cite{DeMicheleLep01,BarbieriEtAl2004} with a Verlet 
algorithm in velocity form and RATTLE algorithm \cite{allentildesley}. Both 
LAMMPS and the in-house software set the time step at $3 \cdot 10^{-3}$, 
yielding an energy drift of about 1 \% in NVE runs. For each state  we averaged 
over at least sixteen different runs ( twenty-four runs at $T=0.6$ due to 
increasing dynamical heterogeneity \cite{EdigerHarrowell12}). This effort was 
needed to reach appreciable statistical accuracy in the evaluation of several 
collective quantities, including the extremely time-consuming evaluation of the 
tiny anisotropies of the monomer random walk, see Sec.\ref{collision}, and the 
VP volume and surface correlation functions, see Sec.\ref{CavRing}.
The equilibration procedure involves runs with time lengths $\Delta t_{eq}$ exceeding at least three times the average reorientation time of the end-end vector \cite{DoiEdwards}. The procedure ensures that  the slowest correlation functions of interest drop at $\Delta t_{eq}$ to a few percent of their maximum value. In order to test the equilibration procedure, we checked if the states under study comply with the universal correlation between the mean square amplitude of the cage rattling and the relaxation in metastable liquids, see Sec.\ref{GenAsp} and refs. 
\cite{OurNatPhys,lepoJCP09,Puosi11,SpecialIssueJCP13,UnivSoftMatter11,DouglasCiceroneSoftMatter12,DouglasStarrPNAS2015,UnivPhilMag11,OttochianLepoJNCS11,CommentSoftMatter13}. Since the correlation is highly sensitive to non-equilibrium effects, the observed perfect agreement, see Fig.\ref{MSDISF} (bottom), provides confidence about the equilibration procedure. It must be pointed out that the present work is interested only in the time window where the structural relaxation is completed. From this respect, given the considerable effort to reach significant accuracy, the  production runs at the lowest temperatures were extended only up to $\sim 10 \, \tau_\alpha$.

Since the model with rigid bonds exhibits weak crystallization resistance, we have paid particular attention to detect any crystallization signature.  The detailed discussion is deferred to Appendix \ref{Appendix2}. We summarize the results: i) no crystalline fraction is revealed in all the systems with semirigid bonds, and the system with rigid bonds at $T=0.9$; ii) the possible crystalline fraction of the system with rigid bonds at $T=0.6$, if present, is so small as to play no role.

\begin{figure}[t]
\begin{center}
\includegraphics[width=0.8\linewidth]{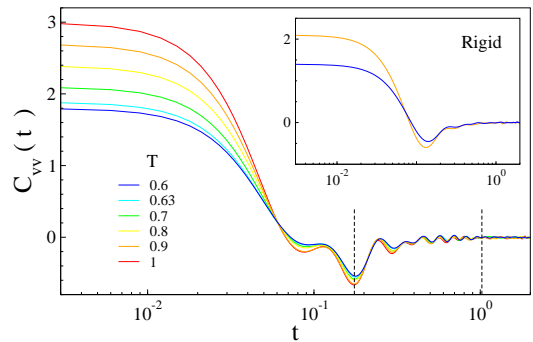} \\
\end{center}
\caption{Velocity correlation function of the monomers at selected temperatures. The left dashed line marks the minimum of the 
correlation function at time $t_m = 0.175$. The position of the right dashed 
line is the time $t^* = 1.023$ defined in Fig.\ref{MSDISF} and marks the end 
of the exploration of the cage by the trapped particle. During the selected 
time 
window ($t_m \lesssim t \lesssim t^*$) the coupling with the solid-like collective motion of the surroundings develops, 
see Sec.\ref{Collision-driven}. The fast oscillations superimposed to the slower decay are largely due to 
the bond vibrations and are mostly suppressed by replacing the semirigid bond with a rigid one (see inset). The residual 
oscillations observed after the minimum in the presence of rigid bonds are ascribed to collective density waves in 
analogy with atomic liquids like rubidium \cite{HansenMcDonaldIIIEd}.}
\label{Cvv}
\end{figure}

\begin{figure}[t]
\begin{center}
\includegraphics[width=0.7\linewidth]{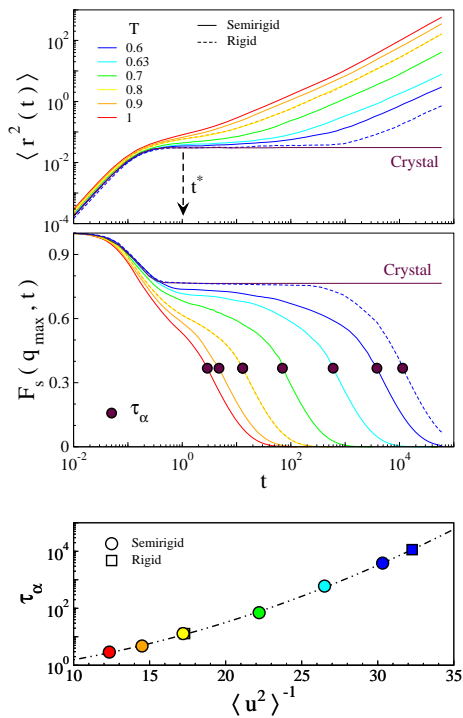} \\
\end{center}
\caption{Monomer dynamics in the molecular liquid. As reference, a crystal 
state having the same density of the other states, obtained by {\it 
spontaneous} crystallization of an {\it equilibrated} liquid made of trimers 
with rigid bonds at $T=0.7$ is also plotted (maroon line). Top: MSD. The knee 
at about $t_m = 0.175$ corresponds to the minimum of the velocity correlation 
function, see Fig.\ref{Cvv}.  Middle: corresponding ISF curves. The long-time 
decay is stretched with stretching parameter $\beta \sim 0.6 $.  Bottom: 
scaling of the non-crystalline states on the universal master curve $\log 
\tau_\alpha$ vs $\langle u^2 \rangle^{-1}$ (dot-dashed line) expressing the 
universal correlation between the fast vibrational dynamics and the slow 
relaxation \cite{OurNatPhys}. $\langle u^2 \rangle$ is the MSD  evaluated at 
the time
$t^*=1.023$ where $\log$ MSD vs. $\log t$ has the inflection point, see top panel. $t^*$ marks the end of the exploration of the cage by the trapped particle.
The expression of the dot-dashed master curve is $\log \tau_\alpha = \alpha + \beta <u^2>^{-1} + \gamma <u^2>^{-2}$ with $\alpha =-0.424(1), \beta  = 2.7(1) \cdot 10^{-2}, \gamma =  3.41(3) \cdot 10^{-3}$ \cite{OurNatPhys}.
}
\label{MSDISF}
\end{figure}

\section{Results and discussion}
\label{resultsdiscussion}

We now present and discuss the results about our trimeric liquid.  The states 
represent a significant set spanning a wide range of relaxation times. Below, 
it will be shown that they exhibit key features of the supercooled liquids, 
e.g. the  stretching of the relaxation 
\cite{OurNatPhys,Ediger00,EdigerHarrowell12}, the presence of dynamical 
heterogeneity \cite{OurNatPhys,NGP_Wolynes,Ediger00,EdigerHarrowell12}, and all 
comply with the universal scaling between the cage rattling and the structural 
relaxation found in several glass-forming systems 
\cite{OurNatPhys,lepoJCP09,Puosi11,SpecialIssueJCP13,UnivSoftMatter11,
DouglasCiceroneSoftMatter12,DouglasStarrPNAS2015,UnivPhilMag11,
OttochianLepoJNCS11,CommentSoftMatter13}. From this respect, we believe that 
the conclusions to be drawn by their analysis are representative of supercooled 
molecular liquids broadly.

\subsection{Transport and relaxation: general aspects}
\label{GenAsp}

The cage effect is well evidenced by the velocity self-correlation function $C_{vv}(t)$ \cite{HansenMcDonaldIIIEd}, 
which is shown in Fig.\ref{Cvv}. Initially, the decay is well accounted for by Eq.\ref{Cvvexpansion}.
Later, a negative region develops due to backscattering from the cage wall leading, on average, to the reversal of the velocity of the particle. A minimum is seen at $t \sim t_m = 0.175$. 
  Superimposed to the slower decay faster oscillations are seen. By replacing the semirigid bond with a rigid one, they 
largely disappear, see Fig.\ref{Cvv} (inset), so that they are ascribed to bond vibrations. Nonetheless, some oscillatory components are still present after the 
minimum. In atomic liquids, e.g. rubidium,  similar components are due to 
sound waves that, owing to the low compressibility, reach wavelengths of a few particle diameters \cite{HansenMcDonaldIIIEd}.

We define the monomer displacement in a time $t$ as:
\begin{equation}
\Delta \mathbf{r}_i (t) = \mathbf{r}_i(t)-\mathbf{r}_i(0)
\label{mondispl} 
\end{equation}
where $\mathbf{r}_i(t)$ is the vector position of the $i$-th monomer at time $t$.
The mean square displacement (MSD) $\langle r^2(t)\rangle$ is expressed as:
\begin{equation}
\langle r^2(t)\rangle = \left\langle \frac{1}{N} \sum_{i=1}^N  \|\Delta 
\mathbf{r}_i (t)\|^2 \right \rangle
\label{Eq:MSD}  
\end{equation}
where brackets denote the ensemble average. In addition to MSD the incoherent, self part of the intermediate scattering function (ISF) is also considered:
\begin{equation}
F_{s}(q,t) = \left\langle \frac{1}{N} \sum_{j=1}^N e^{i{\bf q}\cdot \Delta 
\mathbf{r}_j (t)} \right\rangle
\label{Eq:Fself} 
\end{equation}
ISF was evaluated at $q= q_{max} $, the maximum of the static structure factor ( $7.29 \le q_{max}\le 7.35$  ).

Fig.\ref{MSDISF} shows MSD of the molecular monomers (top) and ISF (middle) curves of the states of interest. At very short times (ballistic regime) MSD increases according to $\langle r^2(t)\rangle \cong (3 k_B T/m) t^2$ and ISF starts to decay. 
The repeated collisions slow the displacement of the tagged monomer, as evinced by the knee of MSD at $ t \sim t_m = 0.175$, i.e. very close to the minimum of the velocity correlation function, see Fig.\ref{Cvv}. 
At later times a quasi-plateau region, also found in ISF, occurs when the temperature is lowered. This signals the increased caging of the particle. Trapping is permanent in the crystalline state so that neither MSD nor ISF decay. In the other states, an inflection point is seen at $t^*=1.023$ in the log-log MSD plot, see Fig.\ref{MSDISF} (top). $t^*$ is state-independent in the present model \cite{OurNatPhys}. The inflection point signals the end of the exploration of the cage by the trapped particle and the subsequent early escapes. The average escape time yields the structural relaxation time $\tau_{\alpha}$, defined by the relation $F_s(q_{max}, \tau_{\alpha}) = e^{-1}$. For $t  > \tau_{\alpha}$ MSD increases more steeply and finally ends up in the diffusive regime, whereas ISF decays to zero as a stretched exponential with stretching parameter $\beta \sim 0.6 $.

Fig.\ref{MSDISF} (bottom) shows that all the states under study (apart from the crystalline state) comply with the universal scaling between the fast vibrational dynamics  and the slow relaxation in glass-forming systems, as expressed by the  master curve between the mean square amplitude of the cage rattling $\langle u^2 \rangle \equiv \langle r^2(t^*)\rangle$ and the structural relaxation time $\tau_\alpha$ \cite{OurNatPhys,lepoJCP09,Puosi11,SpecialIssueJCP13,UnivSoftMatter11,DouglasCiceroneSoftMatter12,DouglasStarrPNAS2015,UnivPhilMag11,OttochianLepoJNCS11,CommentSoftMatter13}.

It is known \cite{OurNatPhys,NGP_Wolynes} that the mean square amplitude of the 
cage rattling $\langle u^2 \rangle$ scales also  the non-gaussian parameter 
$\alpha_2$, a measure of the non-gaussian character of the dynamics, and then of 
its heterogeneous character  ($\alpha_2$ vanishes for gaussian, homogeneous 
dynamics) \cite{Ediger00}. Scaling means that the maximum of the non-gaussian 
parameter $\alpha_2^{max}$ is a universal function of the structural relaxation 
time \cite{OurNatPhys,NGP_Wolynes}.  On this basis, given  the structural 
relaxation time $\tau_\alpha$  of the states under study, we see that they range 
from states with virtually no dynamical heterogeneity, $\alpha_2^{max} \sim 
0.2$, up to states with significant  heterogeneity,  $\alpha_2^{max} \sim 3.4$.

\subsection{Inside the cage}
\label{collision}

\begin{figure}[t]
\begin{center}
\includegraphics[width=0.7\linewidth]{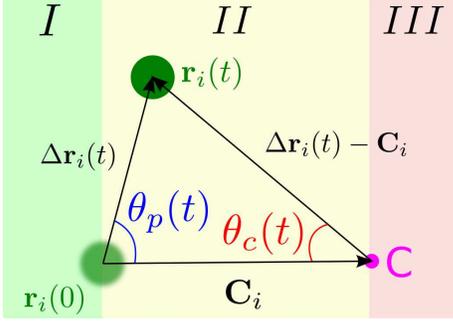} \\
\end{center}
\caption{Quantities of interest to characterize the correlation between the displacement of the  particle (green dot) 
and the cage shape. { The magenta dot is the centroid of the vertices  of the associated Voronoi polyhedron (VP) at the initial time, Eq.\ref{centroid}. The 
highlighted regions are limited by planes perpendicular to  $\mathbf C_i$  and  passing through either the initial 
position of the particle or the centroid.}}
\label{disegnino}
\end{figure}

We now turn our attention on how the cage rattling is affected by the cage shape. 
We correlate the direction of the monomer displacement $ \hat{\bf u}_i (t)$  with the direction of the 
elongation of the VP surrounding the $i$-th particle at the initial time $t=0$. Fig.\ref{disegnino} visualises the 
quantities of interest. 
The direction of the elongation is  defined as \cite{Rahman66}:
\begin{equation}\label{uC}
\hat{\bf u}^C_i = \frac{\mathbf C_i}{ |\mathbf C_i|} 
\end{equation} 
$\mathbf C_i$ is the position of the centroid, the center of mass of the VP vertices, with respect to the position of 
the $i$-th particle:
\begin{equation}
 \mathbf C_i = \frac{1}{N_{v, \, i}}\sum_{j=1}^{N_{v, \, i}} \mathbf v^j_i
 \label{centroid}
\end{equation}
where $N_{v, \, i}$ and  $\mathbf v^j_i$ are the number of vertices and  the position of the VP $j$-th vertex with 
respect to the position of the $i$-th particle, respectively. 

\begin{figure}[t]
\begin{center}
\includegraphics[width=1\linewidth]{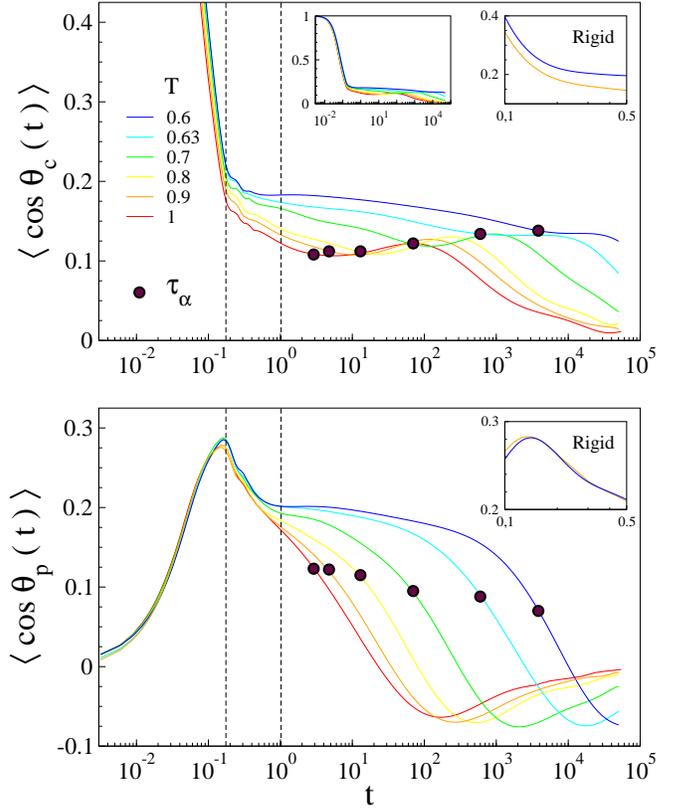} \\
\end{center}
\caption{Local anisotropies of the cage rattling. Top: $\langle\cos\theta_c(t)\rangle$, Eq.\ref{cosTC}, at different 
temperatures. The unlabeled inset shows the complete decay. Bottom:  $\langle\cos\theta_p(t)\rangle$, Eq.\ref{cosTP}. 
The dots mark the structural relaxation time $\tau_\alpha$. The left dashed 
line marks the minimum of the velocity 
correlation function at $t = t_m$, see Fig.\ref{Cvv}. The right dashed line marks the time 
needed by the trapped particle to explore the cage $t^*$, see Fig.\ref{MSDISF}.
Up to $t \sim 0.65$ $\langle\cos\theta_p(t)\rangle$ is nearly 
temperature-independent. The maximum of $\langle\cos\theta_p(t)\rangle$ signals 
 that the particle, as in atomic systems 
\cite{Rahman66},  tends initially to move to the centroid. The insets labelled as "Rigid" plot the anisotropies for 
$T=0.6,0.9$ having replaced the semirigid bond with a rigid one and leaving any other parameter unchanged. They 
evidence that the oscillations in the time window $t_m \lesssim t \lesssim t^*$ of the main panels are due to the 
finite stiffness of the bond.}
\label{coseni}
\end{figure}

In order to investigate the correlation between the displacement of the trapped particle and the shape of the cage, we 
consider the time evolution of two distinct {\it order parameters}, namely the {anisotropy} of the particle 
displacement relative to the centroid with respect to the initial direction of the centroid, see Fig.\ref{disegnino}:
\begin{equation} 
\label{cosTC}
 \langle\cos\theta_c(t)\rangle = \left\langle \frac{1}{N} \sum_{i=1}^N 
\frac{\big[\Delta\mathbf r_i(t) -  \mathbf C_i \big] 
}{\big|\Delta\mathbf r_i(t) -  \mathbf C_i \big|} \cdot 
\big[-\hat{\bf u}^C_i \big] \right\rangle
\end{equation}
and 
the anisotropy of the particle displacement with respect to the initial direction of the 
centroid \cite{Rahman66}, see Fig.\ref{disegnino}:
\begin{equation}
\label{cosTP}
 \langle\cos\theta_p(t)\rangle = \left\langle\frac{1}{N} \sum_{i=1}^N \hat{\bf 
u}_i (t)  \cdot \hat{\bf u}^C_i\right\rangle
\end{equation}
Complete isotropy yields $\langle\cos\theta_i\rangle = 0$ ($i=p,c$). Perfect alignment of $ \hat{\bf u}_i (t)$ 
with respect to $\hat{\bf u}^C_i$ yields $ \langle\cos\theta_p\rangle = 1$ whereas perfect alignment of $\big[\Delta\mathbf 
r_i(t) -  \mathbf C_i \big]$ with respect to $- \hat{\bf u}^C_i$ yields $ \langle\cos\theta_c\rangle = 1$. Furthermore, 
if the monomer displacement is large with respect to $ |\mathbf C_i|$, $\theta_p(t) \simeq \pi - \theta_c(t)$ and  
$\langle\cos\theta_p(t)\rangle \approx -\langle\cos\theta_c(t)\rangle$.  The order parameters defined by Eq.\ref{cosTC} 
and Eq.\ref{cosTP} provide complementary information. By referring to Fig.\ref{disegnino}, positive values of  
$\langle\cos\theta_c(t)\rangle$ signal that the particle is preferentially located in regions I and II, whereas positive 
values of $\langle\cos\theta_p(t)\rangle$ denote preferential location of the particle in regions II and III.

\begin{figure}[t]
\begin{center}
\includegraphics[width=0.7\linewidth]{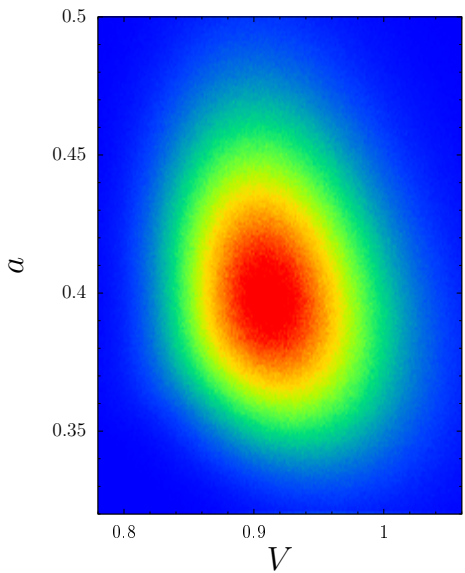} \\
\end{center}
\caption{Correlation plot of the asphericity and the volume of VPs of the melt of trimers at $T=1$ . No correlation is apparent.}
\label{a-S}
\end{figure}

Fig.\ref{coseni} (top) shows detailed plots of $\langle\cos\theta_c(t)\rangle$, Eq.\ref{cosTC}.
At very short times the displacement $\Delta\mathbf r_i(t)$ is small and $\langle\cos\theta_c(t)\rangle \sim 1$. { Then, 
the anisotropy drops in a temperature-independent way up to $t_m$, the time  needed by most particles to reverse their initial velocity, see Fig.\ref{Cvv}}. At later times the decay  slows down and becomes temperature-dependent.  The decay stops at about the structural relaxation time $\tau_\alpha$.
At $ t > \tau_\alpha$ a mild increase of $\langle\cos\theta_c(t)\rangle$ is observed to be followed by a later decay. The description of the non-monotonous relaxation of the order parameter in this viscoelastic regime goes beyond the purposes of the present paper and will be presented  elsewhere.

Fig.\ref{coseni} (bottom) plots $\langle\cos\theta_p(t)\rangle$, Eq.\ref{cosTP},  at different temperatures. At very 
short times the direction of the particle displacement $\hat{\bf u}_i (t)$  is almost isotropic and 
$\langle\cos\theta_p(t)\rangle$ is small. Later, the particle approaches the initial position of the centroid of the VP 
vertices, $\langle\cos\theta_p(t)\rangle$ increases and reaches the maximum at $t_m$, when $C_{vv}(t)$ is at the 
minimum. The initial tendency of the trapped particle to move to the centroid  has been reported 
\cite{Rahman66,DouglasAspehricity} and is clear indication that {\it initially} there is growing correlation between the 
local structure and the particle displacement. However, at later times the correlation decreases,  
up to $t \sim 0.65$ in an almost temperature-independent way. 
For $t \gtrsim 0.65$ the decrease of $\langle\cos\theta_p(t)\rangle$ is slowed down and becomes strongly 
temperature-dependent. The declining anisotropy 
$\langle\cos\theta_p(t)\rangle$, is consistent with our previous finding that 
the influence of both the size and the shape of the cage on the mean square 
displacement is lost within $t^* \sim 1$ 
\cite{VoroBinarieJCP15,VoronoiBarcellonaJNCS14}. For times longer than the 
structural 
relaxation time the escape from the cage reduces the order parameter further, and a negative tail is observed. The tail 
follows by the approximate relation $\langle\cos\theta_p(t)\rangle \approx -\langle\cos\theta_c(t)\rangle$, which holds 
at long times, and the positiveness of $\langle\cos\theta_c(t)\rangle$.

It seems proper to compare the decrease of the two order parameters  in the range $t_m \lesssim t \lesssim t^*$ where structural relaxation is virtually missing  
\cite{OurNatPhys,lepoJCP09,Puosi12}. First, by replacing the semirigid bond with a rigid one, one clarifies that 
the small oscillations which are superimposed to their decay in this time window are due to the finite stiffness of the 
bonds, see Fig.\ref{coseni} (insets).
One also notices that, differently from $\langle\cos\theta_c(t)\rangle$,  $\langle\cos\theta_p(t)\rangle$ 
 is largely temperature-independent if $t \lesssim 0.65$, see Fig.\ref{coseni}. 
This is due to the different character of the two order parameters. { The time-dependence of 
$\langle\cos\theta_p(t)\rangle$ around its maximum at $t_m$ tracks the bounce of the particle with the cage wall. This process is 
nearly {\it temperature-independent}, see Fig.\ref{Cvv}.}  Instead, the anisotropy $\langle\cos\theta_c(t)\rangle$ decreases 
if the population of particles, initially located in regions I and II of Fig.\ref{disegnino}, displaces appreciably to 
region III.  The same process affects $\langle\cos\theta_p(t)\rangle$ less { because region III may be 
reached from region II also with no change of  $\theta_p$}. To reach the region III the particle initially approaches 
the centroid of the VP vertices, located close to the voids between the particles marking the weak spots of the cage. 
The approach to the centroid, and then the decay of 
$\langle\cos\theta_c(t)\rangle$, is limited by the softness of the cage 
\cite{Dyre06,StarrEtAl02,Puosi12,ElasticoEPJE15}, which is {\it 
temperature-dependent}. The elastic response by the cage will be dealt with in 
Sec.\ref{CavRing}.

The above discussion suggests that $\langle\cos\theta_p(t)\rangle$ tracks the monomers being backscattered by the cage wall, whereas $\langle\cos\theta_c(t)\rangle$ is more sensitive to the monomers escaping from the cage. 
This picture is reinforced by observing the changes of the two anisotropies  around $t^*$. We remind that  at $t^*$ early breakouts from the cage start to take place, see Fig.\ref{MSDISF} and Sec.\ref{GenAsp} \cite{OurNatPhys,lepoJCP09}.  $\langle\cos\theta_p(t)\rangle$ has an inflection point at $t^*$ which develops in  $\langle\cos\theta_c(t)\rangle$ only at the lowest temperature, see Fig.\ref{coseni}. The accelerated decay  of $\langle\cos\theta_c(t)\rangle$ around $t^*$ suggests that $\langle\cos\theta_c(t)\rangle$ is more affected by the monomers leaving the cage, since they lose  correlation with the cage geometry, whereas $\langle\cos\theta_p(t)\rangle$ is more sensitive to the trapped particles which keep on being affected by the cage geometry. At low temperature, being the escapes quite rare, the two anisotropies are more similar.

\begin{figure}[t]
\begin{center}
\includegraphics[width=0.7\linewidth]{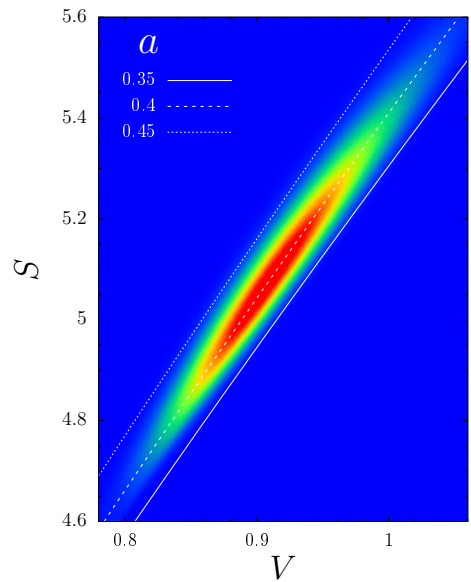} \\
\end{center}
\caption{Correlation plot of the surface and the volume  of VPs of the melt of trimers at $T=1$ (Pearson correlation 
coefficient $r = 0.973$). The lines are Eq.\ref{asphFit} with the indicated values of the asphericity. The best-fit 
value of the asphericity ($a= 0.404$) compares well with the average asphericity of the VPs  ($\langle a \rangle = 
0.405$).}
\label{V-S}
\end{figure}

\begin{figure}[t]
\begin{center}
\includegraphics[width=0.9 \linewidth]{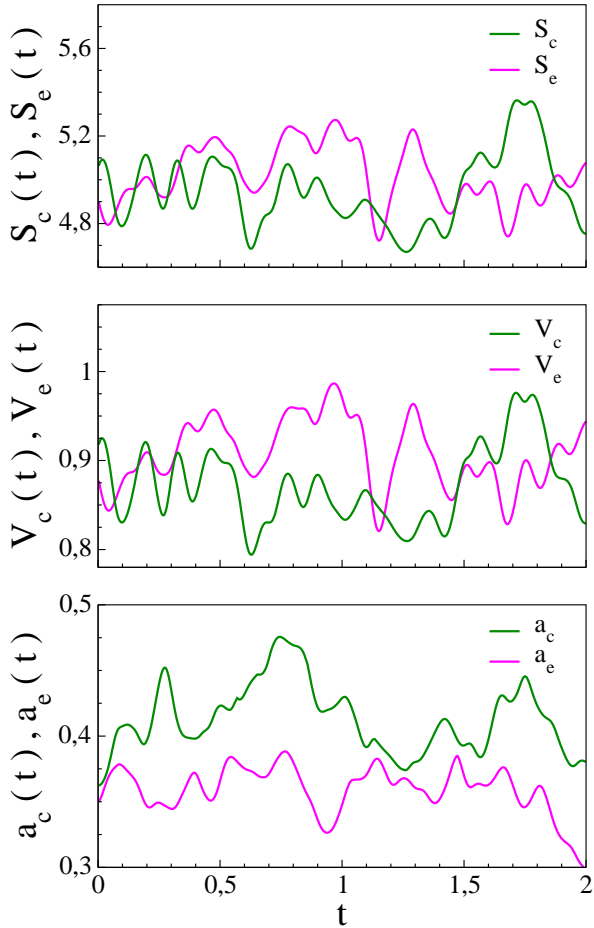} \\
\end{center}
\caption{Selected time-frame of the surface (top), volume (middle) and asphericity (bottom) of the VP surrounding 
particular central (green) and end (magenta) monomers of different trimers of the melt at $T = 1$. Note the extreme 
similarity of the fluctuations of the volume and the surface and the different character of the ones of the 
asphericity.}
\label{V-S2}
\end{figure}

\subsection{Cage border}
\label{CageGeom}

Sec.\ref{collision} discussed how the cage rattling of the trapped monomers is affected by the neighbours. Here, we 
reverse the point of view and investigate how the neighbours are affected by the collisions of the trapped monomer. To 
this aim, we consider the geometry of the cage in terms of the volume $V_i$, the surface $S_i$ and the asphericity $a_i$ 
of the $i$-th VP. The asphericity is defined as:
\begin{equation}
 a_i=\frac{S_i^3}{36\pi V_i^2} -1
 \label{asph}
\end{equation} 
It is non-negative and vanishes for a sphere. For the system under study $ a \sim 0.35-0.45$, namely the VPs are 
moderately non-spherical  (the asphericities of the dodecahedron and the octahedron are $0.325$ and $0.654$, 
respectively )
\cite{VoronoiBarcellonaJNCS14,VoroBinarieJCP15}. { Since each VP includes one particle only, the VP volume  is a measure of the local density.}

\subsubsection{Statics: volume-surface correlations}
\label{VSC}
There are no correlations between  the asphericity and the volume. Fig.\ref{a-S} shows a representative example. No 
correlations are also found between the asphericity and the surface of the VPs (not shown). Instead, Fig.\ref{V-S}  
evidences the strong correlation between the surface and the volume of VPs. 
It follows from the good packing and the subsequent relatively narrow width of the distribution of the asphericity 
\cite{VoronoiBarcellonaJNCS14,VoroBinarieJCP15}.  To show that we recast Eq.\ref{asph} as
\begin{equation}
S = \Big [ 36 \, \pi \, (a+1) \Big ]^{1/3} \, V^{2/3}
 \label{asphFit}
\end{equation} 
Then, we neglect the fluctuations of the asphericity $a$ and treat it as an adjustable parameter to best-fit Eq. 
\ref{asphFit} to the correlation plot. The result is superimposed to the numerical data in Fig.\ref{V-S}. It provides a 
nice fit with best-fit asphericity rather close to the average asphericity of the VPs of the state under consideration. 
The plot also shows that the cloud of data is bounded within the approximate range of the asphericity, $a \sim 
0.35-0.45$, see Fig.\ref{a-S}.

\subsubsection{Dynamics: elastic response}
\label{CavRing}

Volume, surface and asphericity of the $i$-th VP fluctuate around their average values  due to the rearrangement of {\it 
both} the tagged $i$-th particle {\it and} the 
surroundings. 

To give clear impression of the surface-volume correlations we plot in  Fig.\ref{V-S2} a selected time-frame of the 
fluctuations of volume, surface and asphericity of the VP surrounding two specific central and end monomers. The strong 
correlation of the volume and  the surface is quite apparent. Differently, the fluctuations of the asphericity has poor 
resemblance with the ones of the volume and surface.  

To characterize in a quantitative way the fluctuations we define the correlation function:
\begin{equation}
\label{voro}
 C_x(t) = \left\langle \frac{\sum\limits_{i=1}^N \Big[x_i(t)-\langle\langle x 
\rangle\rangle\Big]\Big[x_i(0)-\langle\langle x \rangle\rangle\Big] - 
\delta_x}{\sum\limits_{i=1}^N \Big[x_i(0)-\langle\langle x 
\rangle\rangle\Big]^2-\delta_x}\right\rangle
\end{equation}
with $x = V,S,a$.  $\langle\langle x \rangle\rangle$ denotes the average of $x$ 
over all the $N$ monomers. Eq.\ref{voro} yields $C_x(0) = 1$. To  ensure that 
$C_x(t)$ vanishes at long times we set
\begin{equation}
\label{deltax}
\delta_x =  \frac{N_cN_e}{N}\Big[ \langle\langle x \rangle\rangle_c - 
\langle\langle x \rangle\rangle_e \Big]^2 
 \end{equation}
where $\langle\langle x \rangle\rangle_c$ and $\langle\langle x 
\rangle\rangle_e$ are the average of $x$ restricted to the $N_c$ central and 
the $N_e$ end  monomers, respectively ($N_c+N_e=N$). Eq.\ref{deltax} is derived 
in Appendix \ref{Appendix}.

\begin{figure}[t]
\begin{center}
\includegraphics[width=0.95\linewidth]{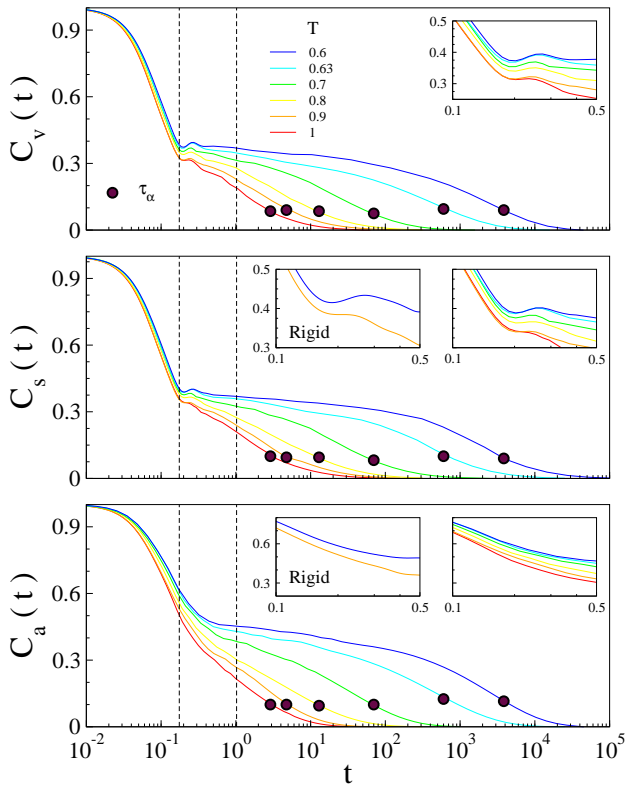} \\
\end{center}
\caption{Correlation functions, Eq.\ref{voro}, of volume (top), surface (middle) and asphericity (bottom) of VPs at 
different
temperatures. The dashed lines mark the region $t_m \le t \le t^*$ as in Fig. \ref{Cvv} and Fig. 
\ref{coseni}. Insets: magnification of the oscillatory behavior at intermediate times. The insets labelled as "Rigid" 
plot the correlation function for $T=0.6,0.9$ having replaced the semirigid  bond with a rigid one and leaving any 
other parameter unchanged. { They evidence that the oscillations are not due to the finite stiffness of the 
bond but to the elastic response of the local structure}.}
\label{voroCorr}
\end{figure}

Fig.\ref{voroCorr} shows $C_x(t)$ with $x= V$ (top panel), $S$ (middle panel) 
and $a$ (bottom panel) at different 
temperatures. As anticipated, there is strong similarity between $C_V(t)$ and $C_S(t)$. { Within the time $t_m$ 
needed by most particles to reverse their initial velocity}, a large part of the correlations of the cage geometry is 
lost. For $t \gtrsim t_m = 0.175$, the decay becomes extremely slow and strongly dependent on the temperature. This parallels 
the time dependence of the order parameter $ \langle\cos\theta_c(t)\rangle$, see Fig.\ref{coseni} (top). At long times, the structural relaxation erases the 
residual correlations of the cage geometry  and $C_x(\tau_\alpha) \simeq 0.1$, 
$x= V,S,a$.

{ We now address the fluctuation correlation in the time window $t_m \lesssim t \lesssim t^*$ where both the velocity correlations, Fig.\ref{Cvv}, and the order parameters, Fig.\ref{coseni}, hint at the elastic response of the cage to the colliding trapped particle. The insets of Fig.\ref{voroCorr} focus on this time interval.}
Correlation oscillations in VP size, but {\it not} in shape, are apparent.  The oscillations are still present if one replaces the 
semirigid bonds by {\it rigid} bonds (see insets labelled as "Rigid" in Fig.\ref{voroCorr}), proving that they are 
not due to bond vibrations. { We interpret the {\it minimum} of $C_V(t)$ and $C_S(t)$ as due to the 
deformation of the local structure following the disordering collision of the central particle with the cage, whereas 
the successive maximum and the later smaller oscillations reflect the elastic response to recover the original 
arrangement. This sort of collective "cage ringing" is not tracked by the asphericity, 
see Fig.\ref{voroCorr} (bottom), confirming the poor correlation between the size and the shape of the cage}. 
It is worth noting that the elastic effects seen via the VP volume and surface are {\it small}, in that the  fluctuations of the VP size and shape are quite limited in size, as seen by Fig.\ref{a-S} and Fig.\ref{V-S}.

\subsection{Beyond the cage border}
\label{Collision-driven}

Sec.\ref{collision} and Sec.\ref{CageGeom}  investigated the cage rattling and the effects on the closest neighbours, respectively. Here, we complete the analysis and extend the range of the neighbours.
It will be now shown that in a time about $t_m = 0.175$, needed by most 
particles to bounce back from the cage wall (Fig.\ref{Cvv}),  extended modes 
involving particles beyond the first shell appear. The modes have solid-like 
character since $\tau_\alpha \gg 1$ \cite{OurNatPhys,lepoJCP09,Puosi12}, i.e. 
they are distinct in nature from the hydrodynamic modes developed by the drag 
force of a moving particle \cite{AlderWainwright70}. We divide the discussion 
in two parts by first describing the onset of extended displacement 
correlations for $t \le t^*$, when the trapped particle completes the cage 
exploration, and then their persistence and decay for $t > t^*$. In particular, 
in Sec. \ref{tstarmeno} we will  compare the influence of the extended modes on 
the moves of the trapped particle in a time $t^*$ with the one of the local 
geometry. Sec.\ref{collision} noted that the influence of the cage shape at 
$t^*$ is smaller than at  $t_m$.

\subsubsection{Onset of the displacement correlations ($t \le t^*$)}
\label{tstarmeno}

To reveal the modes, we characterize the degree of 
correlation between the direction of the displacements performed {\it simultaneously} in the same lapse of time $t$ by two particles {\it initially} spaced by $r_{ij}$ via the space correlation function 
\cite{PuosiLepoJCPCor12,PuosiLepoJCPCor12_Erratum}:
\begin{equation}
\label{Cdir}
C_{\bf{u}}(r,t)= \left\langle\frac{1}{N} \sum_{\substack{i=1}}^{N}   \hat{\bf 
u}_i (t) \cdot  {\bf{U}}_i(r,t)\right\rangle
\end{equation} 
with:
\begin{equation}
\label{CdirAux}
{\bf{U}}_i(r,t)= \frac{1}{N(r)} \sum_{\substack{j=1 \\ i \neq j }}^{N}   \hat{\bf u}_j (t)  
\delta(r-r_{ij}) 
\end{equation} 
where $\hat{\bf u}_i (t)$ and $N(r)$  are the direction of $\Delta \mathbf{r}_i (t)$, Eq.\ref{mondispl}, and the average 
number of particles {\it initially} spaced by $r$, respectively.  If the displacements are perfectly correlated in 
direction, one finds $C_{\bf{u}}(r,t)=1$.  $C_{\bf{u}}(r,t)$ has some formal 
similarity with $ \langle\cos\theta_p(t)\rangle$, Eq.\ref{cosTP}, but the two 
quantities are quite different:
\begin{itemize}
\item  $ \langle\cos\theta_p(t)\rangle$ is a measure of the directional correlation of the displacement performed in a time $t$  by the tagged particle and the axis $\hat{\bf u}^C_i$ set by the {\it initial} cage geometry. 
\item $C_{\bf{u}}(r,t)$ is a measure of the average directional correlation of the simultaneous displacements performed in a time $t$  between the tagged particle and {\it each of the surrounding particles}  at distance $r$. 
\end{itemize}

Fig.\ref{DDcf} (top)  plots the spatial distribution of the correlations for different times $ t$. It is seen that 
if the time is shorter than $t_m = 0.175$,  the time needed by most particles  to reverse their initial velocity 
due to backscattering, the correlations are limited to the bonded 
particles at $r = r_b=b$ and, weakly, the first shell (Fig.\ref{DDcf}, lower panels). { For longer times, the correlation grows in 
both magnitude and spatial extension with characteristic peaks corresponding to the different neighbour shells 
\cite{PuosiLepoJCPCor12,PuosiLepoJCPCor12_Erratum}. These spatial directional correlations have been also observed in simulations on hard spheres and hard disks \cite{DoliwaHeuerPRE00} and experiments on colloids \cite{WeeksJPCM07}. Fig.\ref{DDcf} (lower panels) shows that
the onset of the correlations of both the third and the fourth shells  is delayed of about $0.4$ time units due to the finite propagation speed of the perturbation. It is also seen that the growth of the correlation levels off at times  $\sim t^*$ and is temperature-independent, whereas their magnitude weakly decreases with the temperature. The limited influence of the temperature mirrors the one of the velocity correlation loss in the intermediate range $t_m  \lesssim t \lesssim t^*$. 

Comparing $ \langle\cos\theta_p(t)\rangle$, Fig.\ref{coseni}, and 
$C_{\bf{u}}(r,t)$, Fig.\ref{DDcf},  leads to a clearer picture of how the 
particle in the cage progresses between $t_m$ and $t^*$. Before $t_m$ the cage 
geometry has increasing, even if weak, influence, see Fig.\ref{coseni}. After 
$t_m$, the displacement-displacement correlations increase and extend in space, 
Fig.\ref{DDcf}, with parallel decrease of the influence of the cage geometry, 
Fig.\ref{coseni}. At the completion of the cage exploration at $t^*$, the 
anisotropy of the particle displacement due to local order is small and 
declining, whereas  the displacement-displacement correlations reach their 
maximum. The picture provides an interpretation of the puzzling finding that for 
a given state of a liquid of linear chains (trimers or decamers), the end and 
the central monomers, which have {\it different} distributions of the cage size 
and shape,  have {\it equal} $\langle r^2(t^*) \rangle \equiv \langle u^2 
\rangle$  \cite{VoronoiBarcellonaJNCS14}. On one hand, that finding exposes the 
minor role of the cage geometry. On the other hand, it is well explained by the 
extended displacement-displacement correlations evidenced in Fig.\ref{DDcf}, 
overriding the different local order of the end and the central monomers. The 
leading role of the  extended displacement-displacement correlations in 
setting the monomer moves on the time scale $t^*$ is proven by the fact that  
physical states with {\it identical} 
spatial distributions of the displacement-displacement correlations exhibit  {\it equal} mean square amplitude of the cage rattling $\langle r^2(t^*) \rangle \equiv \langle u^2 \rangle$ 
\cite{PuosiLepoJCPCor12,PuosiLepoJCPCor12_Erratum}.
The coupling between the 
rattling process and extended modes has been indicated 
\cite{Harrowell_NP08,HarrowellJCP09,SastryPRL16,WyartPRL10}.

\begin{figure}[t]
\begin{center}
\includegraphics[width=1\linewidth]{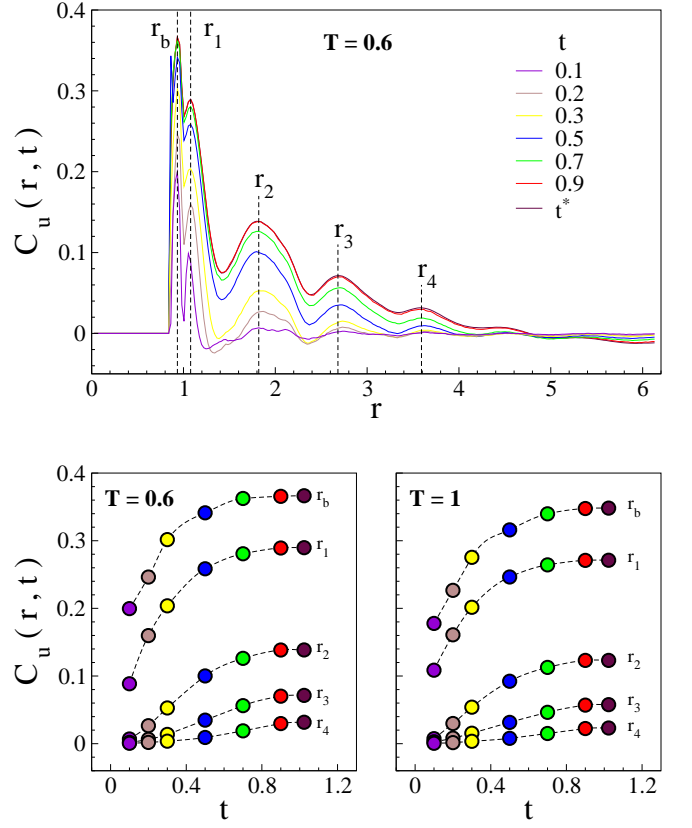} \\
\end{center}
\caption{Top panel: spatial distribution $C_{\bf{u}}(r,t)$ of the correlations between the simultaneous displacements of the central particle 
and  the surrounding particles at distance $r$ for different times $t$ and $T = 
0.6$. The time window starts at $t = 0.1 < t_m = 0.175$ and ends at $t^* = 
1.023$. The peak positions correspond nearly to the ones of 
the radial distribution function \cite{PuosiLepoJCPCor12,PuosiLepoJCPCor12_Erratum}. At constant density they do not 
depend on the temperature. Bottom panels: onset of the correlations at the peak positions at $T=0.6$ (left) and $T=1$ 
(right). $r_b = b$ equals the bond length distance. }
\label{DDcf}
\end{figure}

Displacement correlations have been evidenced in an experimental study of a 
dense colloidal suspension \cite{WeeksJPCM07}. Contact with our simulations is 
allowed by splitting the displacement direction in the transverse and the 
longitudinal components with respect to the direction of the separation vector:
\begin{eqnarray}
 u_m^L&=&\hat{\bf u}_m\cdot\hat{\bf r}_{ij}\\
 u_m^T&=&\hat{\bf u}_m - u_m^L \hat{\bf r}_{ij}
 \end{eqnarray}
where $m = i,j$ and $\hat{\bf r}_{ij} \equiv ({\bf r}_{j} - {\bf r}_{i})/r_{ij}$ refers to the {\it initial} configuration before the displacement occurs. Let us define the related correlation functions as:
\begin{eqnarray}\label{CcompL}
C_L(r, t) = \left\langle\frac{1}{N \; N(r)} \sum_{\substack{i,j=1 \\ i \neq 
j}}^{N}   u_i^L ( t) \cdot  u_j^L ( t)  
\delta(r-r_{ij})\right\rangle \\
\label{CcompT}
C_T(r, t) =\left\langle \frac{1}{N \; N(r)} \sum_{\substack{i,j=1 \\ i \neq 
j}}^{N}   u_i^T ( t) \cdot  u_j^T ( t)  
\delta(r-r_{ij})\right\rangle
\end{eqnarray} 
The longitudinal and the transverse components are related to the total correlation function by:
\begin{equation}
C_{\bf{u}}(r, t)=C_L(r, t)+C_T(r, t).
\end{equation}
Fig.\ref{DDcfcomp} plots $C_L(r, t)$ (top) and $C_T(r, t)$ (middle). The longitudinal correlations increase faster than the transverse ones with increasing  time $ t$. This is seen in  Fig.\ref{DDcfcomp} (bottom) plotting for selected positions the growth function:
\begin{equation}
\label{growth}
 \widetilde C_X(r, t)=\frac{C_X(r, t)-C_X(r,0.1)}{C_X(r,t^*)} \, , \hspace{5mm} X \in\{L,T\}
\end{equation}
The longitudinal correlations have different spatial distribution with respect to the transverse ones. Fig.\ref{DDcfcomp} proves  that the bonded particles ( $r=0.97$) are correlated mostly via their longitudinal displacements. 
Fig.\ref{DDcfcomp} also shows that the oscillatory character of the total displacement correlation function is largely due to the longitudinal component, whereas the transverse component is much less sensitive to the radial density distribution. Other salient features are the negative minimum of $C_L(r, t)$ at $r \simeq 1.4$, corresponding to the first minimum of $g(r)$, and the pronounced maximum of  $C_T(r, t)$ close to the same position. All these hallmarks have been observed  in an experimental study of a dense colloidal suspension \cite{WeeksJPCM07}. This suggests that the key features of the displacement correlations are not strictly affected by the molecular connectivity.

\begin{figure}[t]
\begin{center}
\includegraphics[width=1\linewidth]{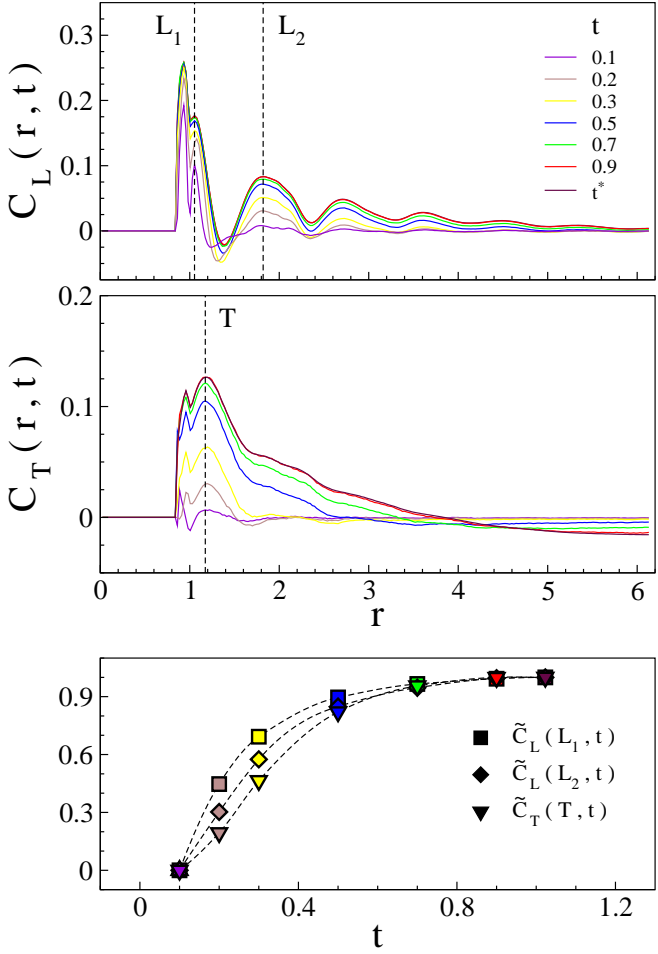} \\
\end{center}
\caption{Spatial distribution of the correlations between the simultaneous longitudinal (top), Eq.\ref{CcompL},  and tranverse (middle), Eq.\ref{CcompT}, components of the displacements of the central particle 
and the surrounding particles at distance $r$ for different times $t$ and $T = 
0.6$. The time window starts at $t = 0.1 < t_m = 0.175$ and ends at $t^* = 
1.023$. The lower panel plots the growth of the correlation, Eq.\ref{growth}, 
at selected positions. The longitudinal correlations increase faster than the 
transverse ones.}
\label{DDcfcomp}
\end{figure}

\begin{figure}[t]
\begin{center}
\includegraphics[width=0.9\linewidth]{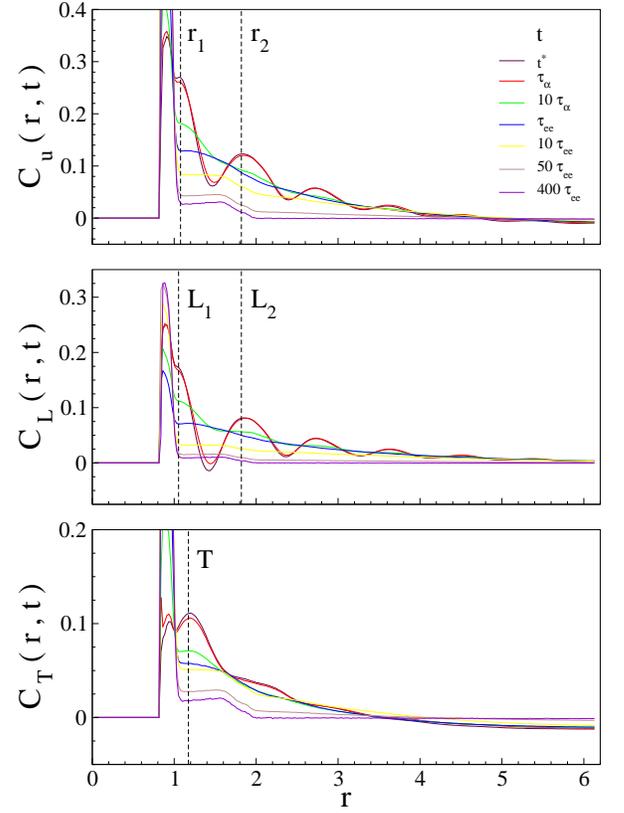} \\
\end{center}
\caption{Spatial distribution of the correlations between the simultaneous overall (top), Eq.\ref{Cdir},  longitudinal (middle), Eq.\ref{CcompL},  and tranverse (bottom), Eq.\ref{CcompT}, components of the displacements of the central particle 
and the surrounding particles at distance $r$ for different times $ t \ge  t^* = 1.023$ and $T = 1$. At the selected temperature $\tau_\alpha = 3$ and the average molecular reorientation time $\tau_{ee} = 126$ \cite{lepoJCP09}.}
\label{morte}
\end{figure}

Even if the full interpretation of the spatial pattern of  $C_L(r, t)$ and $C_T(r, t)$ is deferred to future work, some preliminary remarks may be offered.  {  The negative dip of $C_L(r, t)$ at $r \simeq 1.4$ close to the maximum of $C_T(r, t)$ is consistent with particles approaching , or receding from, each other in a compression/dilation motion while transversely displacing the same way}. The role of quasi-linear arrangements of particles was suggested in regard to the oscillatory behaviour of the longitudinal correlation of the displacements \cite{WeeksJPCM07}. From this respect, evidence of bond-bond alignment, i.e. three monomers in a row,  is reported for the molecular liquid under study \cite{LocalOrderJCP13}. Moreover, in densely packed colloids it is known   that straight paths of $\ell_{SP} $ particles are exponentially distributed as $\sim \exp[-\ell_{SP}/ \bar{\ell}_{SP}]$ with $\bar{\ell}_{SP} \sim 0.73, 0.87$ depending on the sample preparation \cite{SchenkerAsteEuropCerSoc08}. Interestingly, the height of the peaks of $C_{\bf{u}}(r, t)$, due to the longitudinal component, decay exponentially with distance as $\sim \exp[-r/ \xi]$ with $0.7 \le \xi \le 1$ \cite{PuosiLepoJCPCor12,PuosiLepoJCPCor12_Erratum}, suggesting  a connection with the distribution of the length of aligned particles. 

We briefly discuss the weak negative tail observed  in $C_T(r, t)$ at large $r$, Fig.\ref{DDcfcomp} (middle), affecting  $C_{\bf{u}}(r, t)$, Fig.\ref{DDcf}. The tail disappears by increasing the size of the system (not shown) and is due to the momentum conservation requiring that, with fixed center of mass of the system, a displacement of one particle induces correlated counter-displacements on the other ones. The size effect does not affect the longitudinal displacements when averaged over the sphere with radius $r$, see Fig.\ref{DDcfcomp} (top). This may be understood by reminding that the direction of the displacement of the central particle sets the sphere axis. Then, the larger size effect on the transverse displacement follows from the larger weight of the equatorial belt in the average with respect to the polar zones, so that the induced counter-displacements contribute negative terms to $C_T(r, t)$ and negligibly to the average of $C_L(r, t)$.

\begin{figure}[t]
\begin{center}
\includegraphics[width=1 \linewidth]{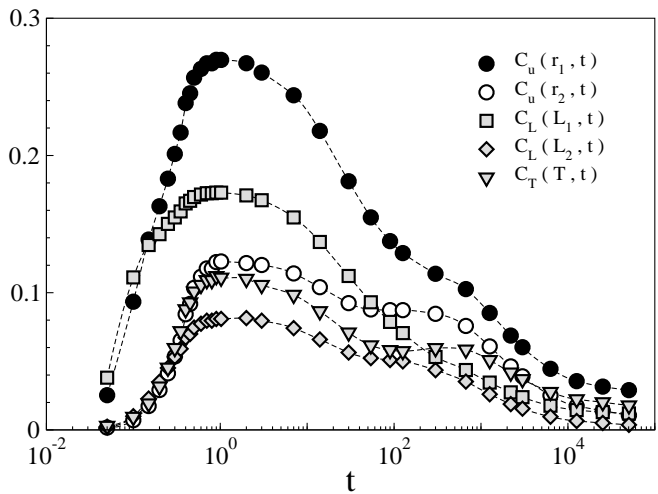} \\
\end{center}
\caption{Time dependence of the overall,  longitudinal and transverse 
correlations at the indicated positions in Fig.\ref{morte}. At the selected 
temperature $\tau_\alpha = 3$ and the average molecular reorientation time 
$\tau_{ee} = 126$ \cite{lepoJCP09}. The time window starts before $t_m = 
0.175$. Note the maximum at about  $t^* = 1.023$.}
\label{summarized}
\end{figure}

\subsubsection{Persistence and decay of the displacement correlations ($t > t^*$)}
\label{tstarpiu}

What happens at the displacement correlations for longer sampling times $ t > t^* = 1.023$  ? We already know that small changes are  observed up to  $ t = \tau_\alpha$ even for sluggish states, thus creating a plateau region on increasing $ t$ \cite{PuosiLepoJCPCor12,PuosiLepoJCPCor12_Erratum}. In this range the presence of quasi-static collective elastic fluctuations \cite{Puosi12} set the magnitude of the direction correlations.  
A  view of the displacement correlations in space for $ t > t^*$ is given in Fig.\ref{morte}. The complete view of the growth, up to $ t \sim t^*$, the plateau region up to $ t \sim \tau_\alpha$, and the following decay is presented in Fig.\ref{summarized}.  Note that, since the decay is quite slow, in order to visualise all the time range, a state with short structural relaxation ($\tau_\alpha = 3$) is considered, thus limiting the persistence of the maximum longitudinal and transverse correlations.
Fig.\ref{morte} shows that the spatial modulation of both the longitudinal and the transverse correlations are averaged for $ t > \tau_\alpha$.  At longer sampling times the magnitude of the correlations decreases further. From this respect, an important time scale is the average molecular reorientation time $\tau_{ee}$,  defined as $C_{ee}(\tau_{ee}) = 1/e$ where $C_{ee}(t)$ is the end-to-end time correlation function  \cite{lepoJCP09}.
The correlations vanish for $r \gtrsim 2$ and $ t \gg \tau_{ee}$. Nonetheless, some permanent correlations are left at shorter distances. In particular, residual correlations due to the intrachain connectivity are present for $1 \lesssim r \lesssim 2$,  together with large correlations between bonded monomers at $r\simeq 0.97$.  Thus, we see that the connectivity of the trimer does not affect the correlation of the displacements of monomers spaced of more than two diameters.
}

\section{Conclusions}
\label{conclusions}

The present paper investigates by MD simulations of a dense molecular liquid 
the key aspects driving the moves of the monomers in the cage of the 
surrounding ones. The aim is clarifying if the displacements are driven by the 
{\it local geometry of the cage} or the {\it solid-like extended modes} excited 
by the monomer-monomer collisions. The main motivations reside in both 
contributing to the intense ongoing research on the relation between the 
vibrational dynamics and the relaxation in glassfoming systems, and improving 
our microscopic understanding  of the universal correlation between the 
relaxation and the mean-square amplitude of the rattling in the cage,  $\langle 
u^2 \rangle$, a quantity related to the Debye-Waller factor.

To discriminate between the roles of the local geometry and the collective extended modes in the single-particle  vibrational dynamics, the cage rattling is examined on different length scales. First,  the anisotropy of the rattling process due to {\it local order}, i.e. the arrangement of the first shell, is characterized by two order parameters sensing the monomers succeeding or failing  to escape from the cage. Then,  the collective response of the surroundings excited by the monomer-monomer collisions is considered. The collective response is initially restricted to the nearest neighbours of the colliding particle by investigating statics and fluctuations of the VP surface, volume and  asphericity.  Then, the {\it long-range} excitation of the farthest neighbours is scrutinised  by searching spatially-extended correlations between the simultaneous fast  displacements of the caged particle and the surroundings. 

Two characteristic times are found: $ t_m = 0.175$ and $t^* \sim 1$. The former is the time when the velocity correlation function reaches the minimum. The latter is the time needed by the trapped particle to explore the cage with mean square rattling amplitude $\langle u^2 \rangle$. One finds that the anisotropy of the random walk  driven by the local order develops up $t_m$, then decreases and becomes small at  $t^* \sim 1$. On the other hand, between $t_m$ and $t^*$ the monomer-monomer collisions excite both the elastic response of the cage and the long-range collective modes of the surroundings in parallel to the decreasing role of the local anisotropies. The longitudinal component of the long-range collective modes has stronger spatial modulation than the transverse one with wavelength of about  the particle diameter, in close resemblance with experimental findings on colloids.

All in all, we conclude that the monomer dynamics at $t^*$, and then $\langle u^2 \rangle$, is largely affected by solid-like extended modes and not the local geometry, in harmony with previous findings  \cite{PuosiLepoJCPCor12,PuosiLepoJCPCor12_Erratum,Puosi12,ElasticoEPJE15,VoroBinarieJCP15,VoronoiBarcellonaJNCS14}, in particular reporting strong, universal correlations with the elasticity \cite{Puosi12,ElasticoEPJE15}, and poor correlation with the size and the shape of the cage  \cite{VoroBinarieJCP15,VoronoiBarcellonaJNCS14}. On a more general ground, our study suggests, in close contact with others \cite{Harrowell_NP08,HarrowellJCP09,SastryPRL16,WyartPRL10}, that the link between the fast dynamics and the slow relaxation is rooted in the presence of modes extending farther than the first shell.

\begin{acknowledgments}
A generous grant of computing time from IT Center, University of Pisa and ${}^{\circledR}$Dell Italia is gratefully acknowledged.
\end{acknowledgments}

\appendix
\section{Derivation of $\delta_x$, Eq.\ref{deltax}}
\label{Appendix}

Let us define the auxiliary quantity:
\begin{eqnarray}
X(t)=\sum_{i=1}^N \Big[x_i(t)-\langle \langle x \rangle 
\rangle\Big]\Big[x_i(0)-\langle \langle x \rangle \rangle\Big] \\
= \sum_{j=1}^{N_c}\Big[\big(x_j(t)-\langle \langle x \rangle 
\rangle\big)\big(x_j(0)-\langle \langle x  \rangle \rangle\big)\Big]  
\label{casino} \\ \nonumber
+ \sum_{h=1}^{N_e}\Big[\big(x_h(t)-\langle  \langle x \rangle 
\rangle\big)\big(x_h(0)-\langle \langle x \rangle \rangle\big)\Big] 
\end{eqnarray}
$\langle \langle x \rangle \rangle$ denotes the average of $x$ over all the $N$ 
monomers. 
Let us define $\langle \langle x \rangle\rangle_c$ and $\langle \langle x 
\rangle \rangle_e$ as the averages of $x$ restricted to the $N_c$ central and 
the $N_e$ end  monomers, respectively ($N_c+N_e=N$). The average $\langle 
\langle x \rangle \rangle$ is related to $\langle \langle x \rangle \rangle_c$ 
and $\langle \langle x \rangle \rangle_e$ by the equation:
\begin{equation}
\label{valoriMedi}
\langle\langle x \rangle \rangle = \frac{N_c}{N} \langle \langle x \rangle 
\rangle_c + \frac{N_e}{N} \langle \langle x \rangle \rangle_e
\end{equation}
In Eq.\ref{casino} we add and subtract   $\langle \langle x  \rangle \rangle_c$ 
to $x_j(t)$,  and  do the same to $x_j(0)$. Also, we add and subtract   $\langle 
\langle x  \rangle \rangle_e$ to $x_h(t)$ and do the same to $x_h(0)$.  If $t 
\to \infty$ the fluctuations of $(x_j(t) - \langle \langle x \rangle \rangle_c)$ 
and the fluctuations of $(x_j(0) - \langle \langle x \rangle \rangle_c)$ have 
both zero average and are uncorrelated. Analogously for the fluctuations of 
$(x_h(t) - \langle \langle x \rangle \rangle_e)$ and the fluctuations of 
$(x_h(0) - \langle \langle x \rangle \rangle_e)$. Then, we yield
\begin{eqnarray}
\label{deltax2}
\delta_x &\equiv&  \lim_{t\to \infty} X(t) \\
\label{deltax3}
&=&N_c \Big[\langle \langle x  \rangle \rangle_c -  \langle \langle x 
\rangle \rangle \Big]^2 + N_e \Big[\langle \langle x \rangle \rangle_e 
- \langle \langle x \rangle \rangle
\Big]^2
 \end{eqnarray}
Plugging Eq.\ref{valoriMedi} into Eq.\ref{deltax3} yields :

\begin{eqnarray}
\nonumber
\delta_x&=&N_c\left[\frac{N_e}{N}\langle \langle x \rangle 
\rangle_c-\frac{N_e}{N} \langle\langle x \rangle
\rangle_e\right]^2+\\
\nonumber
&+& N_e\left[\frac{N_c}{N}\langle \langle x \rangle 
\rangle_c-\frac{N_c}{N}\langle \langle x \rangle\rangle_e\right]^2 \\
&=& \frac{N_cN_e}{N}\Big[ \langle \langle x \rangle\rangle_c - \langle \langle 
x \rangle \rangle_e \Big]^2 \label{pippo2}
\end{eqnarray}
Eq.\ref{pippo2} coincides with Eq.\ref{deltax}.

\section{On the presence of a crystalline fraction}
\label{Appendix2}

Since the model with rigid bonds exhibits weak crystallization resistance, we have paid particular attention to detect any crystallization signature. 
From this respect, we  have monitored some key quantities {\it under equilibration and production  of  both the rigid and semirigid systems}. We summarize the results:

i) the radial distribution function  compares rather well with the one of atomic liquids, apart from the extra-peak due to the bonded monomers.

ii) the pressure and the configurational energy exhibit neither drops nor even slow decreases within 1 \%, during the runs. 

iii) no  global order and even microcrystalline domains are seen by visual inspection of the samples. Note that, if the sample crystallizes, ordering is strikingly visible, e.g. see Fig. 10 of ref.\cite{LocalOrderJCP13}.

iv) no order revealed by Steinhard global order parameters \cite{LocalOrderJCP13}.

v) the mean square displacement always increases steadily with time. In a crystalline sample it levels off, see Fig.\ref{MSDISF} (top).

vi) full decorrelation of the incoherent part of the intermediate scattering function in both equilibration and production runs in all cases except the production runs of the system with rigid bonds at $T=0.6$. In the presence of a solid-like fraction the incoherent part of the intermediate scattering function decays to a finite plateau at long times, see Fig.\ref{MSDISF} (middle).

It is worth noting that the {\it crystalline}  state obtained by {\it spontaneous} crystallization of an {\it equilibrated} liquid made of trimers with rigid bonds at $T=0.7$ considered in Fig.\ref{MSDISF} does not pass {\it any} of the above tests.

The tests iv), v) and vi) are now discussed in detail.

\subsubsection{Test iv): absence of long-range order}
\label{CystFract}

To characterize the degree of {\it global} positional ordering of our samples we resort to the metric $Q_{l,global}$  \cite{Steinhardt83,LocalOrderJCP13}. To this aim, one considers in a given coordinate system the polar and azimuthal angles $\theta({\bf r}_{ij})$ and $\phi({\bf r}_{ij})$ of the vector ${\bf r}_{ij}$ joining the $i$-th central monomer with the $j$-th one belonging to the neighbors within a preset cutoff distance $r^* = 1.2 \; \sigma^* \simeq 1.35$ \cite{Steinhardt83}. The vector ${\bf r}_{ij}$ is usually referred to as a "bond" and has not to be confused with the {\it actual} chemical bonds of the polymeric chain!

\begin{figure}[t]
\begin{center}
\includegraphics[width=0.7\linewidth]{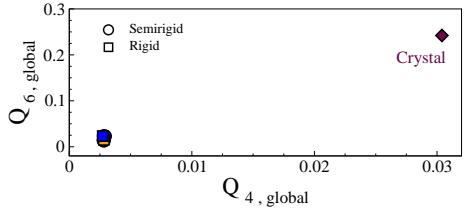} \\
\end{center}
\caption{Global order parameters $Q_{6,global}$ and $Q_{4,global}$. The crystal state is the same considered in Fig.\ref{MSDISF} and is the only state exhibiting global order.}
\label{Steinhard}
\end{figure}

To define a global measure of the order in the system, one calculates the quantity  \cite{Steinhardt83}:
\begin{equation} \label{Qbarlm_global}
\bar{Q}_{lm}^{global}=\frac{1}{N_{b}}\sum_{i=1}^{N}\sum_{j=1}^{n_b(i)}Y_{lm}\Big[ \theta({\bf r}_{ij}),\phi({\bf r}_{ij})\Big]
\end{equation}
where $n_b(i)$ is the number of bonds of $i$-th particle, $N$ is the total number of particles in the system, $Y_{lm}$ denotes a spherical harmonic and $N_b$ is the total number of bonds i.e:
\begin{equation} \label{N_b}
	N_b=\sum_{i=1}^{N} n_b(i) 
\end{equation}
 The global orientational order parameter $Q_{l,global}$ is defined by the rotationally invariant combination:
\begin{equation} \label{Ql_global}
 Q_{l,global}=\left\langle\left [ \frac{4\pi}{(2l+1)} \sum_{m=-l}^{l} 
|\bar{Q}_{lm}^{global}|^2 \right ]^{1/2}\right\rangle
\end{equation}
In the absence of global ordering $Q_{l,global} = 0$ in systems with infinite size. In the presence of long-range crystalline order $Q_{l,global} \neq 0$ , e.g.  $Q_{6,global}  \sim 0.5$  \cite{Steinhardt83,LocalOrderJCP13}, with exact values depending on the kind of order. Fig.\ref{Steinhard} shows $Q_{6,global}$ vs. $Q_{4,global}$ for all the states under investigation. They are compared with one typical crystalline state of the system with rigid bonds. Other crystalline states yield $Q_{4,global}$-$Q_{6,global}$ pairs within the size of the diamond. The non-ideal values of the order parameters of the crystalline state indicate imperfect long-range ordering  \cite{Steinhardt83,LocalOrderJCP13}. We were unable to crystallize the system with semirigid bonds which, however,  having $b = b_{rigid}$, is anticipated to have order parameters similar to the rigid bond case.  Fig.\ref{Steinhard} shows that only the crystalline state  has global, long-range order.

\subsubsection{Tests v) and vi): absence of cristalline fractions}
\label{CystFract2}

Fig.\ref{MSDISF} shows that if the sample crystallizes the changes of both MSD and ISF stop. In principle, the MSD increase in time may be also seen if supercooled liquids and crystalline fraction coexist \cite{DouglasEtAlPNAS2009}. In this heterogeneous case the overall MSD is largely contributed by the mobile phase, since the MSD of the arrested phase levels off rapidly. However, ISF would reach a plateau at long times due to the frozen phase unable to lose the position correlation tracked by ISF. Fig.\ref{MSDISF} shows that ISF vanishes at long times in all states, except one to be discussed below, ruling out the presence of polycrystallinity at the {\it end} of the production runs, i.e. no crystalline regions formed during the equilibration and the subsequent production runs. The ISF of the system with rigid bonds at $T=0.6$  is still non-zero at the end of the production runs (due to our decision to stop the simulation soon after $\tau_\alpha$). To get a rough estimate of the maximum crystalline fraction $\phi_c^{max}$, we identify $\phi_c^{max}$ with the ratio of the residual ISF height, $ISF_r$, with the ISF plateau of the crystal state, $ISF_c$, (weakly dependent on T). We get $\phi_c \le \phi_c^{max} = ISF_r/ ISF_c \simeq 0.09$. We offer arguments to conclude that the possible crystalline fraction  of the system with rigid bonds at $T=0.6$, if present, is much less than the upper limit $\phi_c^{max}$, and, in any case, plays negligible role. In fact: \\
1) the system with rigid bonds at $T=0.6$ passes the tests i), ii), iii), iv), v); \\
2) the same system at $T= 0.9$, where no crystalline fraction is present, provides quite close, or even coincident, results to the ones gathered at $T=0.6$, see the insets of Fig.\ref{Cvv}, Fig.\ref{coseni}, Fig.\ref{voroCorr}. Note also that the results of the systems with rigid and non-rigid bonds are also quite close, see Fig.\ref{Cvv} and Fig.\ref{voroCorr};\\
3) the cage rattling amplitude and the structural relaxation of the system with rigid bonds at $T=0.6$ fulfill the universal scaling between the fast vibrational dynamics  and the slow relaxation in glass-forming systems, see Fig.\ref{MSDISF} (bottom) and Sec.\ref{GenAsp} \cite{OurNatPhys,lepoJCP09,Puosi11,SpecialIssueJCP13,UnivSoftMatter11,DouglasCiceroneSoftMatter12,DouglasStarrPNAS2015,UnivPhilMag11,OttochianLepoJNCS11,CommentSoftMatter13}. The universal scaling is anticipated to fail in semicrystalline materials where the frozen component has finite rattling amplitude but no relaxation. 

\bibliography{biblio.bib}

\end{document}